\documentclass[review,1p]{elsarticle}
\usepackage{graphicx}
\usepackage{amssymb}
\begin{document}
\begin{frontmatter}
%\linespread{1}
%\setlength{\parindent}{0pt}
%\setlength{\parskip}{1ex plus 0.5ex minus 0.2ex}
%\hspace{length}
%\textwidth 6in
%\setlength{\oddsidemargin}{-0.4cm}
%\setlength{\textwidth}{16.8cm}
%\setlength{\topmargin}{-1.25 cm}
%\parskip=4pt
%\textheight 9.5in
%\preprint{YONSEI/2009/10/08}

\title{Little Higgs model with new Z-parity and Dark matter}

\author{C.~S.~Kim}
\ead{cskim@yonsei.ac.kr}
%\address{Department of Physics and IPAP, Yonsei University, Seoul 120-749, Korea}

\author{Jubin Park}
\ead{jbpark@cskim.yonsei.ac.kr}
\address{Department of Physics and IPAP, Yonsei University, Seoul 120-749, Korea}

\begin{abstract}
It has been  known that Little Higgs models with T-parity,
which can give a dark matter candidate,
suffer from the anomalies of given models through the Wess-Zumino-Witten term,
which in turn can violate the T-parity.
Here we thus introduce a new kind of discrete symmetry called Z-parity,
which is a residual gauge symmetry of an additional $\rm{U(1)}$
symmetry of the extended gauge group of the Little Higgs models.
Because Z-parity comes
from the gauge symmetry, this discrete symmetry not only remains unbroken
under the anomalies but also can give the lightest Z-parity particle,
which is stable and becomes a viable dark matter candidate.
%To realize this idea explicitly we choose the mini Little Higgs model as a specific example
%and show that the LXP can be a right-handed Majorana fermion.
We also show that there is an allowed parameter space in the global symmetry breaking scale $f$ versus
the mass of the dark matter, which satisfies the current relic density of the cold dark matter.
\end{abstract}


\end{frontmatter}
%%%%%%%%%%%%%%%%%%%%%%%%%%%%%%%%%%%%%%%%%%%%%%%%%%%%%%%%%%%%%%%%%%%%%%%%
\section{Introduction}
\label{sec:1}
%%%%%%%%%%%%%%%%%%%%%%%%%%%%%%%%%%%%%%%%%%%%%%%%%%%%%%%%%%%%%%%%%%%%%%%%
At one time there had been a wisdom claiming that the supersymmetry is the only candidate
to solve the hierarchy problem without big tuning since the supersymmetry itself gives the beautiful
cancelations between boson and fermion loops and makes these nervous quantum corrections mild.
According to this wisdom, the cancelation is based on the opposite sign between boson and fermion loops in Feynman rules.
However, this kind of cancelation is not unique, that is to say, the possibility
that particles with the same statistics can cancel each other's contribution under quantum corrections may exist.
Actually in the Higgs loops of the minimal supersymmetric standard  model (MSSM) some diagrams among boson loops
cancel each other's loop correction.

Recently one novel and different approach for solving the hierarchy problem was suggested
by Arkani-Hamed, Cohen and Georgi \cite{ArkaniHamed:2001nc}. They
revived the old idea that the Higgs boson is a pseudo-Goldstone boson from a spontaneous symmetry breaking
of approximate global symmetry \cite{Georgi:1974yw},
and made one successful model called the Littlest Higgs model, which can protect the electroweak scale
under quantum loop corrections.
In their first successful model, unlike supersymmetric cases, there are only a few additional new gauge bosons
and a vector-like top quark to cancel their corresponding one-loop corrections of
the partners with the same statistics. This way to cure the harmful quadratic divergences of the squared Higgs mass
is surely more economical and simpler than that of the MSSM which has many more free parameters.
Subsequently, many authors have been inspired by these attractive ideas,
and have built various models~\cite{ArkaniHamed:2002qy},
which are largely classified by the breaking patterns of global symmetry.

As soon as  Little Higgs models had been introduced, these models also received a great attention
from a different point of view besides the stability of the electroweak scale:
Cheng and Low showed that these models can have a discrete symmetry called T-parity like the R-parity
in the MSSM \cite{Cheng:2003ju}, and this T-parity can also give a stable dark matter candidate
by assigning the Standard Model (SM) fields to even and heavy new particles to odd parity, respectively.
Moreover, because the T-parity makes dangerous quantum corrections weak by prohibiting some vertices in Feynman rules,
it can make the Little Higgs models more consistent with severe electroweak precision constraints than before.
However, although the T-parity has good advantages, lately Hill and Hill \cite{Hill:2007zv}
showed that the T-parity can suffer
from anomalies and be broken because of oddness of Wess-Zumino-Witten (WZW) term under T-parity.
This term contains information of anomaly physics of a given theory in the Lagrangian.

In this letter we thus introduce a new discrete $\mathbb{Z}_{2}$ symmetry called Z-parity\footnote{
In our first version of manuscript, we called it the X-parity. However, later we found that the term "X-parity"
has already been used in Ref.~\cite{Freitas:2009jq}. We thanks the authors of \cite{Freitas:2009jq}
for letting us know.}.
This new parity generally comes from a residual discrete gauge symmetry of an additional $\rm{U(1)}$
gauge group~\cite{Ibanez:1991pr}.
In order to realize Z-parity we must simultaneously consider the anomaly-free conditions for the
extended gauge groups  as well as for the relevant Yukawa terms.
As a specific example, we choose the minimal Little Higgs model called Mini Little Higgs model (MLH)~\cite{Bai:2008cf},
only because of the simplicity of the gauge structure $\rm{SU(2)}_{L}\times \rm{U(1)}_{1}\times \rm{U(1)}_{2}$.
We extend MLH and show how to realize Z-parity.
We also show that Z-parity odd particles are right-handed Majorana particles,
and there  exist at least two Z-parity odd particles in the minimally extended model,
therefore, the lightest Z-parity odd particle can be a  viable dark matter candidate.
It is also interesting that, as shown in Table 3, the charge assignments of  Z-parity odd particles
show $\mathbb{Z}_{2}$ symmetries between $\rm{U(1)}_{1}$ and $\rm{U(1)}_{2}$, as well as
among particles in each $\rm{U(1)}$ charge.
Especially, these Z-parity odd particles can interact only with the $Z^{\prime}$ gauge boson and scalar particles.

In addition, we investigate the allowed regions of the parameters
which satisfy the current relic density of the cold dark matter,
% \textbf{from the parameter spaces (global symmetry breaking scale $f$,
% the mass of dark matter $M_{X_{1}}$ and the Yukawa coupling $y_{1}$) of the model},
and find very narrow allowed regions.
It is important that there exists an upper bound of the global symmetry breaking scale, $f \leq 1.1$ TeV,
depending on the dark matter mass (see Fig. 1).
We also show that the mass of top partner $t^{\prime}$ cannot be larger than around 3 TeV
by using the constraint of $f$ value and, therefore, the range of the mass of $t^{\prime}$ is
about $1.4-3$ TeV (see Fig. 2).

\section{Realization of Z-parity in Little Higgs Model}
\label{sec:2}
\textbf{A. Aspects of the Mini Little Higgs model}

MLH is based on the $\rm{SU(2)}_{L}\times \rm{U(1)}_{1} \times \rm{U(1)}_{2}$ gauge symmetry,
and an approximate $\rm{U(3)}$ global symmetry which is broken to $\rm{U(2)}$.
It is important to notice that these $\rm{U(1)}$ charge assignments, which protect the squared mass of
the Higgs particle from radiative quantum corrections of charged weak gauge bosons $W^{\pm}$ and $Z^{0}$ in the SM,
are necessary ingredients in MLH because of the absence of new charged particles.
These heavy charged partners generally cancel the quantum corrections of the $\rm{W^{\pm}, Z^{0}}$
in usual Little Higgs spirit. (Note that there is no additional $\rm{SU(2)}$ gauge group in MLH.)
However, because of the simplicity of the gauge structure we can easily obtain simple anomaly-free constraint
equations for each $\rm{U(1)}$ charge assignment and extend the model to have the new residual
discrete gauge symmetry. We  call this parity Z-parity from now on.

Now we briefly present the lagrangian for the original MLH,
\begin{eqnarray}
\mathcal{L}_{\rm Scalar}^{\rm{~MLH}}&=&
|D_{\mu}\phi |^{2}~~,\\
\mathcal{L}_{\rm Yukawa}^{~\rm{MLH}}&=&y_{1}~\bar{Q}_{L}\tilde{H}t_{R} + \lambda_{b}~\bar{Q}_{L}Hb_{R}
+ \lambda_{e}~\bar{E}_{L}He_{R}+\lambda_{\nu}\bar{E}_{L}\tilde{H}\nu_{R}+ h.c \\
&+&y_{1}\bar{\psi}_{L}St_{R}+y_{2}f\bar{\psi}_{L}\psi_{R} + h.c~~, \nonumber
%&+&\lambda_{X,1}S\left(X_{1}^{T}C^{-1}X_{1}\right)+\lambda_{X,2}S^{\prime}\left(X_{2}^{T}C^{-1}X_{2}\right)h.c~~,
\end{eqnarray}
where $\phi$ is a triplet of the global $\rm{U(3)}$ and decomposed into an $\rm{SU(2)_{L}}$ doublet $H$
and an $\rm{SU(2)_{L}}$ singlet $S$ as $\phi^{T}=(H, S)^{T}$.
The second line describes the usual Yukawa terms plus the coupling term of the right-handed neutrino $\nu_{R}$
for the see-saw mechanism, and the third line presents the mixing between the top quark and heavy top partner $\psi$.
Note that the existence of the same coupling $y_{1}$ for the first terms of the second and third lines,
and the coupling $y_{2}$ for the additional colored vector-like quark $\psi_{L,R}$, are necessary
to cancel the dangerous top loop corrections to the Higgs $h$ field.
(See Ref.~\cite{Bai:2008cf} for details of MLH.)

\begin{table}[h]
\begin{center}
\caption{Examples of anomaly-free $\rm{U(1)}_{1}$ and $\rm{U(1)}_{2}$ charge assignments
for $l_{2}=-h_{2}=-\frac{1}{2}$ (upper two rows) and  $l_{2}=h_{2}=-\frac{1}{2}$ cases (bottom two rows)
where $l_{2}$ is a $z_{2}$ charge for $E_{L}$ and $h_{2}$ is a $z_{2}$ charge for $H$
and we assume that $s_{2}$, which is a $z_{2}$ charge for singlet $S$, is $-\frac{5}{3}$.
In order to get $z_{1}$ charge assignments we use hypercharge relation $z_{1}+z_{2}=\rm{Y}$.}
\begin{tabular}{ccccccccccc}
\hline
\hline
~~ Fields ~~ & ~$Q_{L}$~ & ~$t_{R}$ ~ & ~$b_{R}$~ & ~$E_{L}$~ & ~$\nu_{R}$~ & ~$e_{R}$~ & ~$\psi_{L,R}$~ & ~$H$~ &  ~$S$~ &\\
\hline
\hline
$z_{1}$ & $\frac{1}{6}$ & $\frac{2}{3}$ & -$\frac{1}{3}$ & $-\frac{1}{2}$ & 0 & -1 & $\frac{7}{3}$ & $\frac{1}{2}$ & $\frac{5}{3}$\\
\hline
$z_{2}$ & $\frac{1}{6}$ & $\frac{2}{3}$ & -$\frac{1}{3}$ & $-\frac{1}{2}$ & 0 & -1 & -1 & $\frac{1}{2}$ & -$\frac{5}{3}$\\
\hline
\hline
$z_{1}$ & $\frac{1}{6}$ & $\frac{5}{3}$ & -$\frac{4}{3}$ & $-\frac{1}{2}$ & 1 & -2 & $\frac{10}{3}$ & $\frac{3}{2}$ & $\frac{5}{3}$ \\
\hline
$z_{2}$ & $\frac{1}{6}$ & -$\frac{1}{3}$ & $\frac{2}{3}$ & $-\frac{1}{2}$ & -1 & 0 & $-2$ & $-\frac{1}{2}$ & $-\frac{5}{3}$\\
\hline
\hline
\end{tabular}
\end{center}
\end{table}

With the help of those Yukawa terms, each of which give own constraint equation for each $\rm{U(1)}$ charge,
and anomaly-free constraint equations from the triangular anomalies
under $\rm{SU(2)}_{L}\times \rm{U(1)}_{1} \times \rm{U(1)}_{2}$ gauge group,
we can obtain anomaly-free charge assignments.
We present two examples,  $l_{2}=-h_{2}=-\frac{1}{2}$ case and $l_{2}=h_{2}=-\frac{1}{2}$ case, in Table 1.
Here $l_{2}$ is a $z_{2}$ charge for $E_{L}$ and $h_{2}$ is a $z_{2}$ charge for $H$.
The first case shows an anomaly-free set shown in Ref.~\cite{Bai:2008cf}.
For the details of all Yukawa constraints and anomaly cancelation conditions,
please see the Appendix of this article.

\textbf{B. Realization of Z-parity in the modified Mini Little Higgs model}

Let us start by introducing an additional singlet scalar $S^{\prime}$,
which has the mass around the cutoff energy scale of the model,
and two right-handed Majorana fermions ($X_{1}, X_{2}$)
to achieve the new discrete symmetry as the residual gauge symmetry. These additional particles
and charge assignments under gauge group $\rm{SU(2)}_{L}\times \rm{U(1)}_{1} \times \rm{U(1)}_{2}$
are presented in Table 2.
Note that we cannot yet fix each $\rm{U(1)}$ charge in Table 2 since there are many possible solution sets,
which satisfy necessary constraint equations from Yukawa terms and anomaly-free conditions.
We will show why we need at least two Z-parity odd particles $X_{1}$, $X_{2}$ and
two singlet particles $S$, $S^{\prime}$, through details of these conditions below.

\begin{table}[h]
\begin{center}
\caption{Additional fields and quantum numbers in the modified Mini Little Higgs model (mMLH) with Z-parity.
Each $z_{1}[F]$ and $z_{2}[F]$ denotes each \rm{U(1)} charge for a field $F$.
$S^{\prime}$ is a heavy scalar particle around the cutoff energy scale.
$X_{1}$ and $X_{2}$ are right-handed Majorana fermions.
Y is the Standard Model hypercharge and Q is the electromagnetic charge.}
\begin{tabular}{cccccccc}
\hline
\hline
~~ Fields ~~ & ~~$\rm{SU(3)_{C}}$~~ & ~~$\rm{SU(2)}_{L}$~~ & ~~$\rm{U(1)}_{1}$~~ & ~~$\rm{U(1)}_{2}$~~ & ~~~~Y~~~~ & ~~~~Q~~~~ \\
\hline
$S^{\prime}$ & 1 & 1 & $z_{1}[S^{\prime}]$ & $z_{2}[S^{\prime}]$ & 0 & 0 \\
\hline
$X_{1}$ & 1 & 1 & $z_{1}[X_{1}]$ & $z_{2}[X_{1}]$ & 0 & 0 \\
\hline
$X_{2}$ & 1 & 1 & $z_{1}[X_{2}]$ & $z_{2}[X_{2}]$ & 0 & 0 \\
\hline
\hline
\end{tabular}
\end{center}
\end{table}

In the anomaly structure, $X$ particles can contribute only to $B_{3}$ and $B_{4}$ conditions
(see below) because of darkness (neutral and not weak-charged) of $X$ particles.
Hence, we can easily show that with a single $X$ particle one cannot satisfy these full anomaly-free conditions
due to the complete separation between the anomaly-free MLH sector and the $X$ sector.
Therefore, at least two  $X$ particles are needed to cancel each other's anomaly. Therefore,
 we assume here that the $X$ sector consists of the two right-handed Majorana particles,  $X_{1}$ and $X_{2}$.
Note that introducing of two extra right-handed Majorana fermions
is the minimal way to realize Z-parity in this modified model.

{}From the linear $B_{3}$ condition (of Table 6 in Appendix) each $X_{i}$ must satisfy
\begin{equation}
z_{2}[X_{1}]+z_{2}[X_{2}]=0~.
\end{equation}
Therefore, $z_{2}[X_{1}]=-z_{2}[X_{2}]$.
It is interesting that this charge assignment automatically satisfies other cubic anomaly condition $B_{4}$,
\begin{equation}
z_{2}[X_{1}]^{3}+z_{2}[X_{2}]^{3}=0~.
\end{equation}
However, this charge assignment has inconsistency if, in addition, we consider mass terms of
$X_{1}$ and $X_{2}$  through the Yukawa couplings with only one singlet $S$,
\begin{equation}
\mathcal{L}=\mathcal{L}_{\rm Scalar}^{\rm{~MLH}}+\mathcal{L}_{\rm Yukawa}^{~\rm{MLH}}+\lambda_{X,1}S\left(X_{1}^{T}C^{-1}X_{1}\right)
+\lambda_{X,2}S\left(X_{2}^{T}C^{-1}X_{2}\right)+h.c~,
\end{equation}
because each $X$ has the same Yukawa constraint equation
\begin{equation}
Y_{X}:~~~ z_{2}[S]+2z_{2}[X_{1,2}]=0~.
\end{equation}
Here each $z_{2}[X_{1,2}]$ must have the same charge, $i.e.$ $ - z_2[S]/2=5/6$, which contradicts Eqs. (3,4).
We can thus conclude that the modified Mini Little Higgs model (mMLH) with a single scalar $S$ and Yukawa
couplings cannot satisfy these anomaly free conditions  simultaneously.
Therefore, we have to introduce one more singlet particle $S^{\prime}$,
so that each mass term for $X_{1}$ and $X_{2}$ particles can be generated through separate Yukawa interactions
of the two singlets $S$ and $S^{\prime}$.
It is important to understand that these anomaly cancelation conditions
and additional Yukawa terms are keys to assure a new discrete $\mathbb{Z}_{2}$ symmetry, called Z-parity,
as a residual gauge symmetry. With this simple prescription the Yukawa couplings are rewritten by
\begin{equation}
\mathcal{L}=\mathcal{L}_{\rm Scalar}^{\rm{~MLH}}+\mathcal{L}_{\rm Yukawa}^{~\rm{MLH}}+\lambda_{X,1}S\left(X_{1}^{T}C^{-1}X_{1}\right)
+\lambda_{X,2}S^{\prime}\left(X_{2}^{T}C^{-1}X_{2}\right)+h.c~,
\end{equation}
and  Yukawa constraints are,
\begin{equation}
Y_{X}:~~~ z_{2}[S]+2z_{2}[X_{1}]=0~~,z_{2}[S^{\prime}]+2z_{2}[X_{2}]=0.
\end{equation}
These equations show that if  $z_{2}[S^{\prime}]=-z_{2}[S]$ condition is satisfied,
then all anomaly conditions are automatically fulfilled.
Please note that these Yukawa terms can cause Higgs particle to have a large loop contribution from $X_{1}$ particle
via the coupling between singlet $S$ and $X_{1}$:
The loop effect can revive the previous hierarchy problem.
However, due to the $\frac{1}{f}$ suppression factor from the nonlinear parametrization for singlet $S$,
we can overcome this quadratic divergence for sufficiently large $f$ values.
We will discuss possible parameter regions for $f$,
which satisfy the dark matter relic density, in Sec.~3.

\begin{table}[t]
\begin{center}
\caption{Quantum numbers for the singlet scalar particles $S$ and $S^{\prime}$,
and the right-handed Majorana fermions $X_{1}$ and $X_{2}$ in order to cancel the $\rm{U(1)}^{3}$ and
$\rm{U(1)}-[gravity]^{2}$ anomaly conditions.
Note that all $\rm{U(1)}$ charges are re-scaled by $\frac{3}{5}$
to show simple charge relations between $X_{1}$, $X_{2}$, $S$ and $S^{\prime}$.}
\begin{tabular}{cccccccc}
\hline
\hline
~~ Fields ~~ & ~~$S\rm{U(3)}$~~ & ~~$\rm{SU(2)}_{L}$~~ & ~~$\rm{U(1)}_{1}$~~ & ~~$\rm{U(1)}_{2}$~~ & ~~~~Y~~~~ & ~~~~Q~~~~ \\
\hline
\hline
$X_{1}$ & 1 & 1 & $-\frac{1}{2}$ & $+\frac{1}{2}$ & 0 & 0 \\
\hline
$X_{2}$ & 1 & 1 & $+\frac{1}{2}$ & $-\frac{1}{2}$ & 0 & 0 \\
\hline
$S$ & 1 & 1 & $+1$ & $-1$ & 0 & 0 \\
\hline
$S^{\prime}$ & 1 & 1 & $-1$ & $+1$ & 0 & 0 \\
\hline
\hline
\end{tabular}
\end{center}
\end{table}
%%%%%%%%%%%%%%%%%%%%%%%%%%%%%%%%%%%%%%%%%%%%%

Our final results for mMLH are presented in Table 3. Note that in this table we re-scale
the $\rm{U(1)}$ charges for particles by multiplying by factor $\frac{3}{5}$.
It is now worthwhile to notice that these charge assignments show interesting $\mathbb{Z}_{2}$ symmetry relations
between $\rm{U(1)}_{1}$ and $\rm{U(1)}_{2}$,
as well as between $X$ particles and between scalars $S$ and $S^{\prime}$ for each $\rm{U(1)}$ charge.
It is also remarkable that these particles can interact with $Z^{\prime}$ gauge bosons
under the assignments of $\rm{U(1)}$ charges due to different signs of the charge values.

Before we define Z-parity, we digress to consider the source of the extra singlet
$S^{\prime}$. This singlet $S^{\prime}$ may come from various origins.
For example, this $S^{\prime}$ may be a complete singlet in low energy effective theory
involving new physics at TeV order, or a singlet which breaks an extended gauge group at high energy
beyond the cutoff scale of the model. We briefly assume that the  scalar $S^{\prime}$ particle
comes from the extended gauge group at high energies.
We would like to comment on two possible cases for the extended gauge group:
(I) additional $\rm{U(1)}$ from the symmetry breaking of the grand unified theory
or more enlarged gauge group, or even possibly
(II) the gauge group including left-right symmetry gauge group $\rm{SU(2)}_{L}\times \rm{SU(2)}_{R}$.
(Note that in previous subsection we showed that there exists one interesting charge assignment corresponding
to the $l_{2}=h_{2}=-\frac{1}{2}$ case in Table 1, where only right-handed particles couple to $Z^{\prime}$
gauge boson except for the scalar particles and top partner $t^{\prime}$.)

Finally, we present the long-awaited Z-parity from our anomaly-free charge assignments.
In mMLH the singlet $S$  breaks the original global $\rm{U(3)}$ into $\rm{U(2)}$.
This singlet $S$ also mixes $\rm{U(1)_{1}}$ with $\rm{U(1)_{2}}$, so that we have the hypercharge gauge
symmetry $\rm{U(1)_{Y}}$ and broken $\rm{U(1)^{\prime}}$, which has the heavy $Z^{\prime}$ gauge boson.
After $\rm{U(1)}^{\prime}$ charges are normalized to integers by multiplying by a suitable constant number,
we get the following discrete $\mathbb{Z}_{2}$ parity, which is denoted by $\mathbf{Z}$,
\begin{equation}
\mathbf{Z}~[\textbf{all Fields except~} \it{X}]=\textbf{even}, ~~~~~
\mathbf{Z}~[\it{X}]=\textbf{odd}.
\end{equation}
Two examples are presented in Table 4. As can be seen easily, $X$ particles are the only Z-parity odd particles
under the discrete $\mathbb{Z}_{2}$ symmetry. Therefore, one of these $X$ particles can become a viable dark matter candidate.
\begin{table}[t]
\begin{center}
\caption{Examples of $\rm{U(1)^{\prime}}$ charge assignments and Z-parities ($\bf{Z}$)
for $l_{2}=-h_{2}=-\frac{1}{2}$ (upper two rows) and  $l_{2}=h_{2}=-\frac{1}{2}$ (lower two rows) cases.
Note that all $\rm{U(1)}^{\prime}$ charges are normalized to integers.}
\begin{tabular}{cccccccccccccc}
\hline
\hline
~~ Fields ~~ & ~$Q_{L}$~ & ~$t_{R}$ ~ & ~$b_{R}$~ & ~$E_{L}$~ & ~$\nu_{R}$~ & ~$e_{R}$~ & ~$\psi_{L,R}$~ & ~$H$~ &  ~$S$~ & ~$S^{\prime}$~ & ~$X_{1}$~ & ~$X_{2}$~\\
\hline
\hline
$\rm{U(1)}^{\prime}$ & 0 & 0 & 0 & 0 & 0 & 0 & -4 & 0 & 10 & -10 & -5 & 5 \\
%
% for not normalized charges
% $U_{1}^{\prime}$ & 0 & 0 & 0 & 0 & 0 & 0 & $-\frac{2}{3}$ & 0
% & $\frac{5}{3}$ & $-\frac{5}{3}$ & $-\frac{5}{6}$ & $\frac{5}{6}$ \\
%
\hline
$\bf{Z}$ & 1 & 1 & 1 & 1 & 1 & 1 & 1 & 1 & 1 & 1 & -1 & -1 \\
\hline
\hline
$\rm{U(1)}^{\prime}$ & 0 & 6 & -6 & 0 & 6 & -6 & 4 & 6 & 10 & -10 & -5 & 5 \\
%
% for not normalized charges
% $U_{1}^{\prime}$ & 0 & 1 & -1 & 0 & 1 & -1 & $\frac{2}{3}$ & 1
% & $\frac{5}{3}$ & $-\frac{5}{3}$ & $-\frac{5}{6}$ & $\frac{5}{6}$ \\
%
\hline
$\bf{Z}$ & 1 & 1 & 1 & 1 & 1 & 1 & 1 & 1 & 1 & 1 & -1 & -1 \\
\hline
\hline
\end{tabular}
\end{center}
\end{table}

\section{Dark matter candidate and the relic density}
\label{sec:4}
Let us now consider the dark matter relic density for the lightest $X$ particle.
We have basically four free parameters, $i.e.$
the masses of $X$ particles ($M_{X_{1}}$ and $M_{X_{2}}$), the global symmetry breaking scale ($f$) and
the Yukawa couplings ($y_{1}$ or $y_{2}$). Since $X_{2}$ particle is much
heavier than  $X_{1}$ (note that $X_{2}$ particle has the mass via non-zero VEV of the $S^{\prime}$,
which breaks the high energy gauge group), $X_{1}$ can become the cold dark matter.
We omit the mass parameter $M_{X_{2}}$ from the free parameter space
because this heavy $X_{2}$ particle does not contribute to the annihilation channel of relic density
at the energy scale below a few  TeV.
We choose  $M_{X_{1}}$, $f$ and $y_1$ as free parameters
and investigate allowed regions which can satisfy present relic density, in the interval
$$550~\rm{GeV} \leq M_{X_{1}} \leq 2000~GeV,~ 500~GeV \leq f \leq 2000~GeV,~ 0.8 \leq y_{1} \leq 4.0$$
for fixed $M_{X_{2}}= 10$ TeV.

The current dark matter abundance from WMAP collaboration ~\cite{Dunkley:2008ie} is
$$0.1037 < \Omega_{X_{1}}h^{2}(1\sigma) < 0.1161 \left(0.1099 \pm 0.0062 (1\sigma)\right).$$
To probe the allowed regions, we use
the micrOMEGAs\_2.2~\cite{Belanger:2008sj}, which calculates the relic density as well as direct and indirect
detection rates of the dark matter.
First we show the allowed regions of the parameter space for $f$ and $M_{X_{1}}$ in Fig. 1,
which predicts $f \leq 1.2$ TeV for the allowed range of  $M_{X_{1}} <$ a few TeV.
Here the red (light) and the blue (dark) regions represent allowed regions which satisfy the present relic density
at 95\% (2$\sigma$) and 99\% (3$\sigma$) confidence levels, respectively.
It is interesting that these allowed regions look like step functions
with step-like increase around $500-550$ GeV and $900-1000$ GeV.
In fact, this behavior is closely related to the opening of new annihilation
decay channels, $X_{1}X_{1}\rightarrow hS$ and $X_{1}X_{1}\rightarrow SS$.
In the first step, the opening of the new annihilation channel, $hS$,
entirely dominates annihilation of the dark matter. Namely,
as the relative contribution of the new annihilation channel increases,
the relative contribution of the previously dominant $W^+ W^-$ and $ZZ$ decay channels decrease.
Similarly, in the second step new annihilation decay channel $SS$ begins to open when the
mass of dark matter particle $X_{1}$ approaches about 880 GeV. As the contribution of the new $SS$
decay channel gets larger, the relative contribution of the previous dominant decay channel $hS$ rapidly decreases.
Eventually the contribution of the $SS$ decay channel totally governs  decay modes of the dark matter.

\begin{figure}
\centering
\includegraphics[width=0.7\textwidth]{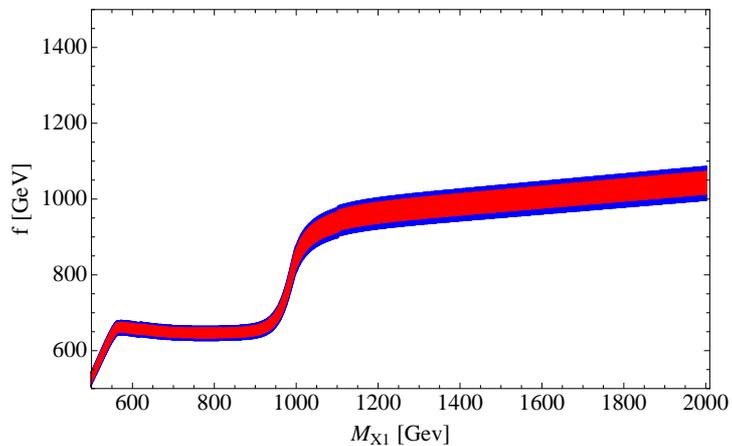}
\caption{The allowed regions which satisfy the current relic density of the dark matter $X_{1}$.
The blue (dark) and red (light) parts correspond to 99\% and 95\% confidence levels, respectively.
The $f$ is a global symmetry breaking scale and $M_{X_{1}}$ is mass of dark matter particle $X_{1}$.
Note that the allowed region is very narrow and there are also upper and lower limits of $f$ along $M_{X_{1}}$.}
\end{figure}

\begin{figure}
\centering
\includegraphics[width=0.7\textwidth]{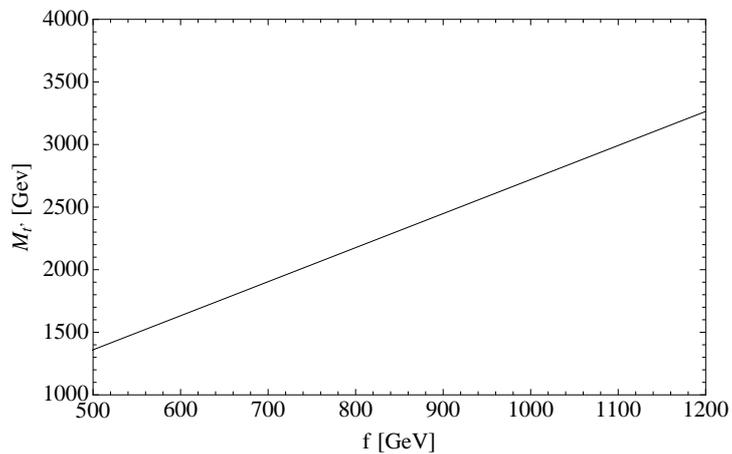}
\caption{The lower bound of the mass of the top partner $t^{\prime}$ as a function of the global symmetry breaking scale $f$ from the mixing between top and top quark partner.}
\end{figure}

Because the top quark contribution in the relic density is so negligible, we may expect
that $y_{1}$ parameter can also be omitted from the four free parameters.
However, the value of $y_{1}$ is important to relate
the mass of top with top partner particles. From the Yukawa terms (See Eq. (2))
we can calculate the relevant masses by
\begin{equation}
m_{t}=\frac{y_{1}y_{2}}{\sqrt{y_{1}^{2}+y_{2}^{2}}}~h~,~~ m_{t^{\prime}}=\sqrt{y_{1}^{2}+y_{2}^{2}}~f,
\end{equation}
where $m_{t}$ is mass of the top quark and $m_{t^{\prime}}$ is mass of the top partner.
In Fig. 2 we show the lower bound of the mass of top partner $t^{\prime}$
as a function of symmetry breaking scale $f$.
It is interesting that the mass range of the top partner $t^{\prime}$
is about $1.4-3$ TeV for the allowed values of $f \leq 1.2$ TeV, which is estimated from Fig. 1.

%%%%%%%%%%%%%%%%%%%%%%%%%%%%%%%%%%%%%%%%%%%%%%%%%%%%%%%%%%%%%%%%%%%

We note that recently there have been interesting observations
from PAMELA \cite{Adriani:2008zr} and ATIC \cite{atic} experiments,
 which support a very striking energy source of energetic electrons
and positrons within $\sim$1 kpc away from the Solar System.
At present this source has not yet been fully identified. The source may be the annihilation of dark matter particles
or an astrophysical object.
Some papers show that these objects may be pulsars which produce a bump in the spectrum \cite{Profumo:2008ms},
and recent results from the Fermi Telescope \cite{Abdo:2009zk} suggest that the ATIC result
may not be a signal of the annihilation of the dark matter particles.
In mMLH with Z-parity, these $X$ particles seem to be  difficult to have selective annihilations
to charged leptons in order to produce the excess of electrons and positrons.
Consequently, the $X$ particles are not the sources of the ATIC and PAMELA observations.

Finally, we remark that the modified Mini Little Higgs Model has two interesting implications:

(I) Various Little Higgs models have extended gauge symmetries,
$e.g.$, some models have the extended hypercharge gauge symmetry like $\rm{U(1)}_{1}\times \rm{U(1)}_{2}$
before the global symmetry breaking.
Therefore, we can apply our Z-parity to these kinds of general Little Higgs models,
and we can infer that there can exist stable particles other than usual T-parity stable particles.

(II) Another interesting aspect is a possibility to explain the neutrino masses, cold dark matter (CDM)
and hierarchy problems simultaneously.
There are papers which explain the CDM and neutrino mass simultaneously but not
the hierarchy problem~\cite{Kubo:2006rm}.
However, mMLH can not achieve above aspect because of the minimal structure of the scalar sector
of MLH.

In conclusion, we showed how to realize Z-parity in the modified Mini Little Higgs model,
and found that Z-parity can give a new kind of stable and neutral dark matter candidate,
the right-handed Majorana particle.
Some additional fields $S^{\prime}$ and $X_{2}$ particles were also introduced
to accomplish the anomaly-free conditions and to obtain simultaneously the Yukawa terms leading to a residual gauge symmetry in the model. Moreover, we also found that the dark matter $X_{1}$ can satisfy
the present dark matter relic density, and there exist allowed regions in the free parameter set of
the model. %which satisfy the relic density.
Finally, the mass range of the top partner $t^{\prime}$ was also deduced
for the allowed values of the global symmetry breaking scale $f$.

\newpage

\centerline{\bf ACKNOWLEDGEMENTS}
\noindent C.S.K. was supported in part by Basic Science Research Program through the NRF of Korea
funded by MOEST (2009-0088395) and  in part by KOSEF through the Joint Research Program (F01-2009-000-10031-0).
JBP thanks S. K. Kang for valuable discussions and encouragement about this paper
and H. S. Lee for his kind and critical opinions.
\\

\appendix{APPENDIX.}
\\

In this appendix, we briefly introduce the particles and corresponding quantum numbers of MLH,
and then discuss the Yukawa terms and anomalies in order to help understand of the
main body. Subsequently, we show how to obtain the anomaly-free charge
assignments in MLH, and present anomaly-free solution sets.

\begin{table}[h]
\begin{center}
\caption{Fields and quantum numbers in the Mini Little Higgs model.
Each $z_{i}[F]$  denotes $\rm{U(1)}$ charge for a field $F$.
Here $\psi_{L,R}$ is a vector-like colored top partner,
and $S$ is a scalar particle that breaks the global $\rm{U(3)}$ symmetry.
Y is the Standard Model hypercharge and Q is the electromagnetic charge.}
\begin{tabular}{cccccccc}
\hline
\hline
~~ Fields ~~ & ~~$S\rm{U(3)_{C}}$~~ & ~~$\rm{SU(2)}_{L}$~~ & ~~$\rm{U(1)}_{1}$~~ & ~~$\rm{U(1)}_{2}$~~ & ~~~~Y~~~~ & ~~~~Q~~~~ \\
\hline
$Q_{L}=\left( \frac{t_{L}}{b_{L}} \right)$ & 3 & 2 & $z_{1}[Q]$ & $z_{2}[Q]$ & $\frac {1}{6}$ & $\frac{2}{3},-\frac{1}{3}$ \\
\hline
$t_{R}$ & 3 & 1 & $z_{1}[t_{R}]$ & $z_{2}[t_{R}]$ & $\frac{2}{3}$ & $\frac{2}{3}$ \\
\hline
$b_{R}$ & 3 & 1 & $z_{1}[b_{R}]$ & $z_{2}[b_{R}]$ & $-\frac{1}{3}$ & $-\frac{1}{3}$ \\
\hline
$E_{L}=\left( \frac{\nu_{L}}{e_{L}} \right)$ & 1 & 2 & $z_{1}[E_{L}]$ & $z_{2}[E_{L}]$ & $-\frac {1}{2}$ & 0,-1 \\
\hline
$\nu_{R}$ & 1 & 1 & $z_{1}[\nu]$ & $z_{2}[\nu]$ &  0 & 0 \\
\hline
$e_{R}$ & 1 & 1 & $z_{1}[e_{R}]$ & $z_{2}[e_{R}]$ & -1 & -1 \\
\hline
$\psi_{L}$ & 3 & 1 & $z_{1}[\psi_{L}]$ & $z_{2}[\psi_{R}]$ & $\frac{2}{3}$ & $\frac{2}{3}$ \\
\hline
$\psi_{R}$ & 3 & 1 & $z_{1}[\psi_{R}]$ & $z_{2}[\psi_{R}]$ & $\frac{2}{3}$ & $\frac{2}{3}$ \\
\hline
$H=\left( \frac{h+}{h^{0}} \right)$ & 1 & 2 & $z_{1}[H]$ & $z_{2}[H]$ & $\frac {1}{2}$ & 1,0 \\
\hline
$S$ & 1 & 1 & $z_{1}[S]$ & $z_{2}[S]$ & 0 & 0 \\
%\hline
%$S^{\prime}$ & 1 & 1 & $z_{1}[S^{\prime}]$ & $z_{2}[S^{\prime}]$ & 0 & 0 \\
%\hline
%$X_{1}$ & 1 & 1 & $z_{1}[X_{1}]$ & $z_{2}[X_{1}]$ & 0 & 0 \\
%\hline
%$X_{2}$ & 1 & 1 & $z_{1}[X_{2}]$ & $z_{2}[X_{2}]$ & 0 & 0 \\
\hline
\hline
\end{tabular}
\end{center}
\end{table}
First, the particles and corresponding quantum numbers under
$\rm{SU(3)}_{C}\times \rm{SU(2)}_{L}\times \rm{U(1)}_{1}\times \rm{U(1)}_{2}$ are presented in Table 5.
The first column is the symbol of particle. The second and third columns are the $\rm{SU(3)}_{C}$
and $\rm{SU(2)}_{L}$ charges for color and chiral weak interaction, respectively.
Next two columns are the $\rm{U(1)}$ charges of the corresponding field denoted by $z_{i}[F]$,
where $i$ is the index of each $\rm{U(1)}$ gauge group.
Note that we have not yet assigned the $\rm{U(1)_{1}}$ and $\rm{U(1)_{2}}$ charges
because we have to discuss various constraints from Yukawa terms and anomalies of the given model.
These $\rm{U(1)}$ charge assignments protect MLH
from the radiative quantum corrections up to $5-10$ TeV scale.

\begin{table}[h]
\begin{center}
\caption{Anomalies and anomaly cancelation conditions. Here $A_{j}$'s denote
the anomalies involving $\rm{U(1)}_{1}$ only.
Similarly, $B_{j}$'s denote the anomalies involving $\rm{U(1)}_{2}$ only.
$C_{j}$'s are used as the mixed anomalies involving $\rm{U(1)}_{1}$ and $\rm{U(1)}_{2}$.}
\begin{tabular}{ccc}
\hline
\hline
Identifier & Anomaly & Anomaly cancelation condition\\
\hline
$A_{1}$ & $\rm{U(1)}_{1}-[\rm{SU(2)}_{L}]^{2}$ & tr$[z_{1}\tau^{a}\tau^{b}]=\frac{1}{2}\delta^{ab}\sum_{f_{L}}~z_{1,f_{L}}$=0~ (doublet fermions only) \\
$A_{2}$ & $\rm{U(1)}_{1}-[S\rm{U(3)}_{C}]^{2}$ & tr$[z_{1}t^{a}t^{b}]=\frac{1}{4}\delta^{ab}\sum_{q}~z_{1,q}$=0~(color triplet fermions only) \\
$A_{3}$ & $\rm{U(1)}_{1}-[$gravity$]^{2}$ & tr$[z_{1}]=\sum_{f}~z_{1,f}$=0~ ($\rm{U(1)}_{1}$-charged fermion only) \\
$A_{4}$ & $[\rm{U(1)}_{1}]^{3}$ & tr$[z_{1}^{3}]=\sum_{f}~z_{1,f}^{3}$=0~ ($\rm{U(1)}_{1}$-charged fermion only) \\
\hline
$B_{1}$ & $\rm{U(1)}_{2}-[\rm{SU(2)}_{L}]^{2}$ & tr$[z_{2}\tau^{a}\tau^{b}]=\frac{1}{2}\delta^{ab}\sum_{f_{L}}~z_{2,f_{L}}$=0~ (doublet fermions only) \\
$B_{2}$ & $\rm{U(1)}_{2}-[S\rm{U(3)}_{C}]^{2}$ & tr$[z_{2}t^{a}t^{b}]=\frac{1}{4}\delta^{ab}\sum_{q}~z_{2,q}$=0~ (color triplet fermions only) \\
$B_{3}$ & $\rm{U(1)}_{2}-[$gravity$]^{2}$ & tr$[z_{2}]=\sum_{f}~z_{2,f}$=0~ ($\rm{U(1)}_{2}$-charged fermion only) \\
$B_{4}$ & $[\rm{U(1)}_{2}]^{3}$ & tr$[z_{2}^{3}]=\sum_{f}~z_{2,f}^{3}$=0~ ($\rm{U(1)}_{2}$-charged fermion only) \\
\hline
$C_{1}$ &~~~ $[\rm{U(1)}_{1}]^{2}-\rm{U(1)}_{2}$ & ~~~~~~~tr$[z_{1}^{2}z_{2}]=\sum_{f}~z_{1,f}^{2}~z_{2,f}$=0  \\
$C_{2}$ &~~~ $\rm{U(1)_{1}}-[\rm{U(1)}_{2}]^{2}$ & ~~~~~~~tr$[z_{1}z_{2}^{2}]=\sum_{f}~z_{1,f}~z_{2,f}^{2}$=0  \\
\hline
\hline
\end{tabular}
\end{center}
\end{table}

We summarize all important anomalies in Table 6,
and divide all these anomalies into three classes ($A$, $B$ and $C$).
Here $A_{j}$'s are the anomalies involving $\rm{U(1)}_{1}$ only.
Similarly $B_{j}$'s denote the anomalies involving $\rm{U(1)}_{2}$ only.
The mixed anomalies between these $\rm{U(1)}_{1}$ and $\rm{U(1)}_{2}$ are denoted by $C_{j}$'s.

Now we consider the solution sets of the $\rm{U(1)}$ charges.
In MLH  we have 18 unknown parameters such as
$z_{i=1,2}[F]$, where  $F=\{Q_{L}, t_{R}, b_{R}, E_{L}, \nu_{R}, e_{R}, \psi_{L,R}, H, S\}$.
There are also $10=2 \times 5$ constraints $:$ 5 Yukawa constraints ($Y_{U}, Y_{D}, Y_{E}, Y_{N}, Y_{S}$)
per each $\rm{U(1)}$, where  subscripts denote up, down, electron, right-handed neutrino
and the singlet $S$, respectively.
Moreover, there are 9 hypercharge relations  and
10 anomaly  conditions ($A_{1-4},B_{1-4},C_{1,2}$).
Therefore, the number of constraints is larger than that of unknown parameters. However, all these constraints are not completely independent.
For instance, anomaly conditions $A_{2,3}$ and $B_{2,3}$ give the constraint equations
which are the same as Yukawa constraints given above.
In order to find the solutions of the valid charges for $\rm{U(1)}_{1}$ and $\rm{U(1)}_{2}$ efficiently,
it would be better to use hypercharge relations first.
These 9 hypercharge relations reduce our original 18 unknown parameters to 9 unknown parameters,
and simplify two ($C_{1}$ and $C_{2}$) anomaly cancelation conditions
by adding and subtracting these two conditions.
Then, we can show that these two nonlinear equations become simpler,
\begin{equation}
\hat{C_{1}}:~~\sum_{f}Y_{f}z_{2,f}^2=0~,
\end{equation}
\begin{equation}
\hat{C_{2}}:~~\sum_{f}Y_{f}^{2}z_{2,f}=0~.
\end{equation}
Since 5 Yukawa constraints and $B_{1}$ give a total of 6 linear constraints
for 9 unknown parameters for the $\rm{U(1)}_{2}$ sector, three parameters
still remain to be unknown.
Therefore, general solution sets of $\rm{U(1)}_{2}$ are given by
\begin{equation}
\left( \begin{array}{c} z_{2}[Q_{L}]\\ z_{2}[t_{R}] \\ z_{2}[b_{R}] \\
z_{2}[E_{L}] \\ z_{2}[\nu_{R}] \\ z_{2}[e_{R}] \\
z_{2}[\psi_{L,R}] \\ z_{2}[H] \\ z_{2}[S] \\ \end{array}\right)
=\frac{l_{2}}{3}\left( \begin{array}{c} -1 \\ -1 \\ -1 \\ 3 \\ 3 \\ 3 \\ -1 \\ 0 \\0 \\ \end{array} \right)
+h_{2} \left( \begin{array}{c} 0 \\ 1 \\ -1 \\ 0 \\ 1 \\ -1 \\ 1 \\ 1 \\ 0\\ \end{array} \right)
+s_{2} \left( \begin{array}{c} 0 \\ 0 \\ 0 \\ 0 \\ 0 \\ 0 \\ 1 \\ 0 \\ 1 \\ \end{array} \right),
\end{equation}
where $l_{2}=z_{2}[E_{L}]$, $h_{2}=z_{2}[H]$ and $s_{2}=z_{2}[S]$.
It is interesting that these general solution sets automatically satisfy remaining constraints,
quadratic $\hat{C_{1}}$, $\hat{C_{2}}$ and cubic $B_{4}$. We present here two example assignments
for $l_{2}=h_{2}=\frac{1}{2}$ and $l_{2}=-h_{2}=\frac{1}{2}$ cases in Table 1.
%In Ref.~\cite{Bai:2008cf} the $l_{2}=h_{2}=\frac{1}{2}$ case is chosen as one example,
%and this assignment is obviously anomaly-free because each $z_{i}$ charge has identical value of
%the hypercharge value($\rm{Y}$) except $\psi$ and $S$.
%It is also important that because the SM particles have the same $z_{i}$ charges,
%the $\rm{Y}_{Z^{\prime}}$ (charge of $Z^{\prime}$) values are all zero for the SM particles
%under an assumption that each $\rm{U(1)}$ gauge group has the same coupling constant,
%$g_{1}^{\prime}=g_{2}^{\prime}$.
%The SM particles, thus, cannot interact with the $Z^{\prime}$ gauge boson.
%On the other hand the $l_{2}=-h_{2}=\frac{1}{2}$ case shows
%interesting couplings between $Z^{\prime}$ and right-handed SM particles because of the
%different signs of the same charge values of $z_{1}$ and $z_{2}$.
%%%%%%%%%%%%%%%%%%%%%%%%%%%%%%%%%%%%%%%%%%%%%%%%%%%%%%%%%%%%
\\

\end{document}